\documentclass{article}

\input{tcilatex}

\begin{document}

\title{{\Large Higher Order Thermal Corrections to Photon Self Energy}}
\author{}
\maketitle

\begin{center}
Mahnaz Q. Haseeb$^{1}$and\ Samina S. Masood$^{2}$

$^{1}$\textit{Physics Department, COMSATS Institute of Information
Technology, Islamabad, Pakistan,}

$^{2}$\textit{Department of Physics, University of Houston Clear Lake,
Houston TX 77058}

\bigskip

\textbf{Abstract}
\end{center}

We investigate temperature behavior up to two loop level in QED in the
background heat bath using real time formalism. The thermal correction to
the coupling constant in QED at low temperature are presented up to the two
loop level. It is observed that the removal of singularities at the two loop
level is only possible if the integrations of loop momenta are done in a
specific order. Some of the possible applications of the results are also
indicated.

\section{Motivations\qquad \qquad }

It is now well-known that the temperatures were too high to ignore thermal
contributions to particle processes in the very early universe. Quantum
field theory at finite temperature is being studied for more than 50 years
to investigate these effects. Techniques for calculations in many body
systems were originally developed in condensed matter physics [1 ] and were
\ used to describe a large ensemble of multi-particle interactions in a
thermal background. The techniques of high temperature field theory are
being used to understand the phase transitions in cosmology and were adopted
in quantum field theories in late seventies. They are extensively used to
describe the phase transitions due to symmetry breaking after the Hot Big
Bang and in tracing the history of the early universe. The methods can be
utilized not only in particle physics but also to investigate various
aspects of nuclear matter and plasma physics.

Some of our results can be applied to study the physical systems where
temperatures are high enough to include the background contributions. In
most of the cases the density effects are also nonignorable and these
techniques can easily be generalized to incorporate them.

The early universe provides a very good example for studying hot plasmas.
Right after the big bang, the universe was impossibly hot and dense. It
rapidly expanded and cooled. Around $t\leq 100$ sec$,$ $T=10^{9}K$ \ [2],
the universe was cool enough for neutrons and protons to combine to form
Deuterium, then Helium and traces of Lithium (primordial nucleosynthesis).
For the next few $10^{5}$ years it was still too hot for electrons to form
atoms. The universe was filled with hot plasma of electrons and nuclei,
bathed in photons constantly interacting with both, such as the interior of
a star.

Similar type of nucleosynthesis occurs in the stellar cores at very high
temperatures. Specifically, the cores of supermassive stars like supernova
and neutron stars are very hot and dense. When the high mass stars exhaust
their Helium fuel they have enough gravitational energy to heat up to $%
6\times 10^{8}K$ $\ $[3]. Cores of neutron stars, red giants and white
dwarfs are composed of extremely dense plasmas ($\rho =10^{6}$ $-10^{15\text{
}}g/cm^{3})$. The neutrinos and axions emission rates in these stars require
the incorporation of thermal background [4]. A tremendous amount of energy
is released in a supernova. The only supernova in modern time, visible to
the naked eye, was detected on Feb. 23, 1987 and is known as $SN1987A.$ It
emitted more than $10^{10}$ times as much visible light as the Sun for over
one month and temperatures as high as $2\times 10^{11}K$ $\ $were reached.

There is a series of different type of fusion reactions in stars leading to
luminous supergiants. When helium fusion ceases in the core, gravitational
compression increases the core's temperature above $6\times 10^{8}K$ at
which carbon can fuse into neon and magnesium. As the core reaches $%
1.5\times 10^{8}K,$ oxygen begins fusing into silicon, phosphorous, sulfur,
and others. At $2.7\times 10^{8}K,$ silicon begins fusing into iron. This
process essentially stops with the creation of iron and a catastrophic
implosion of the entire star initiates.

The quark gluon plasma is the form of matter at transition temperatures $%
T_{c}=100-200$ $MeV$. The hot and dense environment in quark gluon plasma
and the studies of its prospective reproduction in nucleus-nucleus
collisions require the methods of thermal field theory for detailed
investigations. With the increased feasibility of creation of quark gluon
plasma in heavy ion collisions, the methods developed in this theory got
their specific relevance in non-perturbative QCD at finite temperature as
well.

Some of the methods of quantum field theory are very useful to understand
the material properties and the transport theories. These mathematical
techniques can thus be used to study some of the applications of condensed
matter physics [5]. In the next section we describe the finite temperature
corrections to vacuum polarization tensor. \ Sections III and IV give the
one loop and higher order corrections respectively. Finally Sec. V comprises
of discussion of the results.\qquad

\section{Finite Temperature Effects\qquad \qquad \qquad \qquad \qquad }

The main idea of finite temperature field theory is to use the approach of
the usual quantum field theory. Matsubara [6] was first to develop thermal
field theory by incorporating a purely imaginary time variable in the
evolution operator. In Euclidean space the covariance breaks and time is
included as an imaginary parameter. The imaginary time domain is finite and
periodic because of which the energy integrations are converted into
summations over the discrete Matsubara frequencies. The presence of discrete
energies alongwith the particle distribution functions destroys the
covariance of the theory.

The important contributions by Schwinger [7], Mills [8], and Keldysh [9] led
to the development of a formalism based upon the choice of a contour in the
complex plane. This is called the real time formalism. In the real-time
formalism, an analytical continuation of the energies along with Wick's
rotation restores covariance in Minkowski space at the expense of Lorentz
invariance. The breaking of Lorentz invariance leads to the non-commutative
nature of the gauge theories [10]. The covariance is incorporated through
the 4-component velocity of the background heat bath $u^{\mu }=(1,0,0,0).$
In a heat bath the particles are in constant interaction with the thermal
surroundings. Implementing these interactions is straightforward as is done
in vacuum field theory. The temperature is included through the statistical
distribution functions of the particles.

Umezawa and coworkers [11] independently worked on a different approach
called Thermofield dynamics that also gives the same results. In this
formalism, the propagators are taken in the form of $2\times 2$ matrices.
Field theory at finite temperature is renormalizable since the presence of
the Boltzmann factor in the thermal corrections cuts off any ultraviolet
divergence. Choosing the suitable counter terms, as in vacuum, can eliminate
them. The infrared divergences are inherent in almost all perturbation
theories, whether at zero or finite temperature. KLN theorem [12]
demonstrates that singularities appearing at intermediate stages of the
calculation cancel in the final state physical result.

Quantum electrodynamics (QED) is the simplest and most successful gauge
theory. The behavior of QED at finite temperatures serves as a model for the
determination of background effects in other physical theories - the
electroweak theories as well as quantum chromodynamics. In the real time
formalism, the tree level fermion propagator in Feynman gauge in momentum
space is [13]

\begin{equation}
S_{\beta }(p)=(\NEG{p}-m){\Large [}\frac{i}{p^{2}-m^{2}+i\varepsilon }-2\pi
\delta (p^{2}-m^{2})n_{F}(E_{p}){\Large ]}\ ,
\end{equation}%
where

\begin{equation}
n_{F}(E_{p})=\frac{1}{e^{\beta (p.u)}+1}\ ,
\end{equation}%
\bigskip is the Fermi-Dirac distribution function with $\beta =\frac{1}{T}$.
The boson propagator is

\begin{equation}
D_{\beta }^{_{\mu \nu }}(p)={\Large [}\frac{i}{k^{2}+i\varepsilon }-2\pi
\delta (k^{2})n_{B}(k){\Large ]},
\end{equation}%
with

\begin{equation}
n_{B}(E_{k})=\frac{1}{e^{\beta (k.u)}-1}.
\end{equation}%
\bigskip

\section{One Loop Corrections}

At the one loop level, Feynman diagrams are calculated in the usual way by
substituting these propagators in place of the ones in vacuum. The Lorentz
invariance breaking and conserving terms remain separate at the one loop
level since the propagators comprise of temperature dependent (hot) terms
added to temperature independent (cold) terms. This effect has been studied
in detail and established at the one-loop level [14]. The renormalization of
QED\ in this scheme [15] has already been checked in detail at the one loop
level for all temperatures and chemical potentials.

The thermal background effects are incorporated through the radiative
corrections. In finite temperature electrodynamics, electric fields are
screened due to the interaction of the photon with the thermal background of
charged particles. The physical processes take place in a heat bath
comprising of hot particles and antiparticles instead of vacuum. The exact
state of all these background particles is unknown since they continually
fluctuate between real and virtual configurations. The net statistical
effects of the background fermions and bosons enter in the theory through
the fermion and boson distributions respectively.

The electric permittivity and the magnetic susceptibility of the medium are
modified by incorporating the thermal background effects. At low
temperatures, \textit{i.e}., $T<<m_{e}$ ($m_{e}$ is the electron mass), the
hot fermions contribution in background is suppressed and only the hot
photons contribute from the background heat bath. The vacuum polarization
tensor in order $\alpha $ does not acquire any hot corrections from the
photons in the heat bath. This is because of the absence of self-interaction
of photons in QED.

The thermal mass is generated radiatively. The mass shift that enters into
physical quantities acts as a kinematical cut off, in the production rate of
light weakly coupled particles from the heat bath. The effective mass
corresponds to the fact that in the heat bath, the propagation of particles
is altered by their continuous interactions with the medium.\bigskip

\section{Higher Order Corrections}

The higher order loop corrections are required to get predictions on
perturbative behavior at finite temperature. At the higher-loop level, the
loop integrals involve a combination of cold and hot terms which appear due
to the overlapping propagator terms in the matrix element. In such
situations, specific techniques are needed, even at the two loop level, to
solve them. Higher loops get analytically even more complicated. In the hot
terms there appear overlapping divergent terms. The removal of such
divergences is already shown at the two loop level [16] for electron self
energy.

We restrict ourselves to the low-temperatures, for simplicity, to prove the
renormalizability of QED at the two-loop level through the order by order
cancellation of singularities. The $\delta (0)$ type pinch singularities
also appear in Minkowski space. The problem of pinch singularities has been
resolved in thermofield dynamics by doubling the degrees of freedom [11].
The particle propagators become 2$\times $2 matrices whose 1-1 elements
correspond to the usual thermal propagators. We used an alternative
technique to get rid of these $\delta (0)$ type pinch singularities through
the identity [17]

\begin{equation}
i\pi \left[ \delta (k^{2})\right] ^{2}=-\frac{\delta (k^{2})}{%
k^{2}+i\varepsilon }-\frac{1}{2}\delta ^{\prime }(k^{2}).
\end{equation}

In this technique, the results depend on the order of doing the hot and cold
integrations. As an illustration of the difference of the order of
integration, we compare the results of one of the terms. Consider the
singular terms when the hot loop in Fig. (1a) is evaluated before the cold
one, we simply get%
\begin{equation}
g^{\mu \nu }\Pi _{\mu \nu }^{a}(p,T)=\frac{\alpha ^{2}T^{2}}{3}{\Large (}1-%
\frac{2}{\varepsilon }{\Large )},
\end{equation}%
\ \ whereas, in the same term, the evaluation of the hot loop after the cold
one gives 
\begin{equation}
g^{\mu \nu }\Pi _{\mu \nu }^{a}(p,T)=-\frac{\alpha ^{2}}{\pi ^{2}}{\Large [}%
\frac{4\pi ^{2}T^{2}}{3\varepsilon }-\frac{\pi ^{2}T^{2}}{5}-\frac{2T^{3}}{%
5m^{2}}\zeta \left( 3\right) {\Large (}3|\mathbf{p}|+\frac{49}{3}p_{_{0}}%
{\Large )}+\frac{52p_{0}^{3}T^{4}}{5m^{4}|\mathbf{p}|}{\Large ]}.
\end{equation}

The justification of this specific order is the fact that the temperature
dependent part corresponds to the contribution of real background particles
on mass-shell and incorporates thermal equilibrium. The breaking of Lorentz
invariance changes these conditions for the cold integrals. We have checked
that the renormalization can only be proven with the preferred order of
integrations, i.e., if covariant hot integrals are evaluated before the cold
ones. At the higher loop level the vacuum polarization contribution is non
zero, even at low temperature. The calculations are simplified if the
temperature dependent integrations are performed before the temperature
independent ones. The temperature independent loops can then be integrated
using the standard techniques of Feynman parametrization and dimensional
regularization as in vacuum [18]. At the two loop level, the vertex type
corrections to the virtual electron in Fig. (1b) vanish and the self energy
type corrections to the electron loop in Fig. (1a) contribute.

\FRAME{ftbpF}{4.4815in}{1.0205in}{0pt}{}{}{Figure}{\special{language
"Scientific Word";type "GRAPHIC";maintain-aspect-ratio TRUE;display
"USEDEF";valid_file "T";width 4.4815in;height 1.0205in;depth
0pt;original-width 8.2226in;original-height 1.8516in;cropleft "0";croptop
"1";cropright "1";cropbottom "0";tempfilename
'JA9UIP02.bmp';tempfile-properties "XPR";}}The presence of the statistical
effects of photons modifies the vacuum polarization and hence the electron
charge. This leads to the changes in the electromagnetic properties of the
hot medium even at low temperatures. We have calculated the electromagnetic
coupling constant in QED in the photon background up to the second order.
The longitudinal and the transverse components of the vacuum polarization
tensor are\qquad 
\begin{equation}
\Pi _{L}(p,T)=-\frac{p^{2}}{|\mathbf{p}|^{2}}u^{\mu }u^{\nu }\Pi _{\mu \nu
}(p,T)=\frac{2\alpha ^{2}T^{2}p^{2}}{3|\mathbf{p}|^{2}}\left( 1+\frac{%
p_{0}^{2}}{2m^{2}}\right) ,
\end{equation}%
and%
\begin{equation}
\Pi _{T}(p,T)=-\frac{1}{2}[\Pi _{L}(p,T)-g^{\mu \nu }\Pi _{\mu \nu
}^{a}(p,T)]=\frac{\alpha ^{2}T^{2}}{3}{\Huge [}\frac{1}{2}-\frac{p^{2}}{|%
\mathbf{p}|^{2}}\left( 1+\frac{p_{0}^{2}}{2m^{2}}\right) {\Huge ]}.
\end{equation}%
respectively. These components of the vacuum polarization tensor can then be
used to determine the electromagnetic properties of a medium with hot
photons.

\section{Results}

The electron mass and charge renormalization has been obtained from the two
loop self-energy for electrons and photons respectively in Ref. [19]. The
renormalizability of the self mass of electron is proven through the order
by order cancelation of singularities at both loop levels. The charge
renormalization constant of QED, the electron charge renormalization up to
the order $\alpha 
{{}^2}%
$ can be expressed as

\begin{equation}
Z_{3}=1-\frac{\alpha }{3\pi \varepsilon }+\frac{\alpha ^{2}T}{6m^{2}}^{2}.
\end{equation}

The second term in the above equation, the first order contribution is
calculated in detail in Ref. [14 ]. The corresponding value of the QED\
coupling constant is now

\begin{equation}
\alpha _{R}=\alpha (T=0)\left( 1-\frac{\alpha }{3\pi \varepsilon }+\frac{%
\alpha ^{2}T^{2}}{6m^{2}}\right) .
\end{equation}

The results [16] are an explicit proof of renormalizability of QED up to the
two-loop level. They also estimate the temperature dependent modification in
the electromagnetic properties of a medium. This helps to evaluate the decay
rates and the scattering cross-sections of particles in such a medium. These
results can be applied to check the abundance of light elements in
primordial nucleosynthesis, baryogenesis and leptogenesis. If the background
magnetic fields are also incorporated then one can look for applications to
neutron stars, supernovae, red giants, and white dwarfs.

\bigskip

{\Large References}

\begin{enumerate}
\item A. L. Fetter and J. L. Walecka, \textit{Quantum Theory of Many
Particle Systems}, (McGraw-Hill, New York, 1971).

\item E. Kolb, M. S. Turner, \textit{Early Universe}, D. A. Dicus, \textit{%
et.al}., Phys. Rev. \textbf{D26}, 2694 (1982); J. L. Primack and M. Sher,
Nucl. Phys. B209, 372 (1982); \textit{ibid} \textbf{B222}, 517(E) (1983);

http://www.damtp.cam.ac.uk/user/gr/public/bb\_history.html.

\item Bradley W. Carroll and Dale A. Ostlie, \textit{Introduction to Modern
Astrophysics}, (Benjamin Cummings, 1996).

\item Stuart Shapiro and Saul Teukolsky, \textit{Black Holes, White Dwarfs,
and Neutron Stars}, (John Wiley and Sons, 1983).

\item D. A. Kirzhnits, \textit{Field Theoretical Methods in Many Body
Systems }(Permagon, Oxford, 1967); E. M.\textit{\ }Lifshitz and L. P.
Pitaevskii, \textit{Course on Theoretical Physics- Physical Kinetics}
(Pergamon Press, New York).

\item T. Matsubara, Prog. Theor. Phys. \textbf{14} (1955) 351.

\item J. Schwinger, J. Math. Phys. \textbf{2 }(1961) 407]

\item R. Mills, \textit{Propagators for Many Particle Systems} (Gordon and
Breach, New York, 1969).

\item L. V. Keldysh, Sov. Phys. \textbf{20} (1964) 1018.

\item C. Brouder, A. Frabetti \textbf{hep-ph/0011161} and F. T. Brandt,
Ashok Das, J. Frenkel, Phys. Rev. \textbf{D65} (2002) 085017.

\item H. Umezawa, H. Matsumoto and M. Tachiki, \textit{Thermo Field Dynamics
and Condensed States} (North Holland, Amsterdam, 1982); A. Das, \textit{%
Finite Temperature Field Theory }(World Scientific, Singapore, 1997) and
several other references.

\item T. Kinoshita, J. Math. Phys. \textbf{3} (1962) 650.

\item J. F. Donoghue and B. R. Holstein, Phys. Rev. \textbf{D28} (1983) 340;
E. Braaten and R. D. Pisarski, Nucl. Phys. \textbf{B337} (1990) 567; R.
Kobes, Phys. Rev. \textbf{D42} (1990) 562; L. Dolan and R. Jackiw, Phys.
Rev. \textbf{D9} (1974) 3320; P. Landsman and Ch G. Weert, Phys. Rep. 
\textbf{145} (1987) 141 and the references therein.

\item K. Ahmed and Samina Saleem (Masood), Phys. Rev. \textbf{D35} (1987)
1861; \textit{ibid}\ (1987) 4020 and several other papers refered therein;
J. F. Donoghue, B. R. Holstein, and R. W. Robinett, Ann. Phys. (N.Y.) 
\textbf{164}, 233 (1985); K. Ahmed and Samina S. Masood, Ann. Phys. (N.Y.) 
\textbf{207} (1991) 460; Samina Masood, Phys. Rev. \textbf{D44}, (1991) 3943
and references therein.

\item Samina S. Masood, Phys. Rev. \textbf{D47} (1993) 648; \textit{ibid}\
Phys. Rev. \textbf{D36} (1987) 2602.

\item Mahnaz Qader (Haseeb), Samina S. Masood, and K. Ahmed, Phys. Rev. 
\textbf{D44} (1991) 3322; \textit{ibid }Phys. Rev. \textbf{D46}, (1992) 5633
and references therein.

\item L. R. Mohan, Phys. Rev. \textbf{D14 }(1976) 2670.

\item See for example: C. Itzykson and J. B. Zuber, \textit{Quantum Field
Theory }(McGraw- Hill Inc., 1990).

\item Samina S. Masood and Mahnaz Q. Haseeb \textbf{hep-ph/0406079; }Samina
S. Masood and Mahnaz Q. Haseeb \textbf{hep-ph/0612136.}
\end{enumerate}

\end{document}